\documentclass{article}

    \usepackage[preprint]{neurips_2023}

\usepackage[utf8]{inputenc} 
\usepackage[T1]{fontenc}    
\usepackage{hyperref}       
\usepackage{url}            
\usepackage{booktabs}       
\usepackage{amsfonts}       
\usepackage{nicefrac}       
\usepackage{microtype}      
\usepackage{xcolor}         
\usepackage[square,numbers]{natbib}
\usepackage{amsmath}
\usepackage[pdftex]{graphicx} 
\usepackage{caption}
\usepackage{subcaption}

\newtheorem{theorem}{Theorem}[section]
\newtheorem{lemma}[theorem]{Lemma}

\title{A parametric distribution for exact post-selection inference with data carving}

\author{
  Erik Drysdale\thanks{http://www.erikdrysdale.com}
}

\newcommand{\indep}{\perp \!\!\! \perp}

\begin{document}

\maketitle

\begin{abstract}
    Post-selection inference (PoSI) is a statistical technique for obtaining valid confidence intervals and p-values when hypothesis generation and testing use the same source of data. PoSI can be used on a range of popular algorithms including the Lasso. Data carving is a variant of PoSI in which a portion of held out data is combined with the hypothesis generating data at inference time. While data carving has attractive theoretical and empirical properties, existing approaches rely on computationally expensive MCMC methods to carry out inference. This paper's key contribution is to show that pivotal quantities can be constructed for the data carving procedure based on a known parametric distribution. Specifically, when the selection event is characterized by a set of polyhedral constraints on a Gaussian response, data carving will follow the sum of a normal and a truncated normal (SNTN), which is a variant of the truncated bivariate normal distribution. The main impact of this insight is that obtaining exact inference for data carving can be made computationally trivial, since the CDF of the SNTN distribution can be found using the CDF of a standard bivariate normal. A python package \href{https://pypi.org/project/sntn/}{\texttt{sntn}} has been released to further facilitate the adoption of data carving with PoSI.
\end{abstract}

\section{Introduction \label{sec:intro1}}
Post-selection inference (PoSI) is a powerful framework for obtaining statistically valid results in the presence of adaptive hypothesis testing \cite{fithian2017, taylor2016, lee2013, tian2015, loftus2015, markovic2017, lockhart2013, tibs2014, lee2014}. The familiar ``most powerful tests'' and confidence intervals from the classical testing framework rely on having null hypotheses specified independently of the data \cite{lehmann1950, lehmann1955}. In modern data science applications, practitioners will often use the same data to both fit models and conduct inferences. 

Historically, exploratory data analysis was seen as hypothesis generating \cite{tukey2019}, with valid inference only possible by collecting futher data or using sample splitting \cite{cox1975}. The PoSI literature has shown that many sparsity inducing linear regression-based algorithms frequently used by data scientists, such as the Lasso or forward stepwise regression, have hypothesis generating events which amount to a set of polyhedral constraints on the response vector \cite{taylor2015}. These constraints permit the construction of exact statistical tests in finite samples under certain parametric assumptions. For example, the non-zero coefficients selected by the Lasso procedure can be used as a basis for running a traditional regression model, and these coefficients will follow a truncated normal distribution under the null hypothesis. The PoSI methods discussed in this paper are different in nature to the de-biasing methods which are based on asymptotics and stronger regularity conditions  \cite{javanmard2014, vandegeer2014, zhang2014}.

There is a fundamental trade-off between splitting data into hypothesis generating and testing sets. As more data is put towards the hypothesis generating set, more true signals are likely to be tested, but this will reduce the probability of rejecting erroneous null hypotheses. In this paper, PoSI will refer to using 100\% of the data for both selection and inference, while sample splitting will refer to using a portion of data for selection, and a different portion for inference. Data carving, in contrast, will refer to using one portion of data for both testing and inference, and the other portion for only inference. Compared to sample splitting, data carving strictly dominates in terms of power \cite{fithian2017}, and produces confidence intervals whose length is bounded \cite{kivaranovic2020}.

Because data carving combines data from two different distributions, existing papers have used computationally costly sampling methods like MCMC to carry out inference \cite{buhlmann2021, tian2016, fithian2017}. This paper shows that, contrary to existing beliefs, data carving has a well-defined parametric distribution when the response is a Gaussian and the conditioning event is characterized by a set of polyhedral constraints. This distribution, the sum of a normal and a truncated normal (SNTN), though not often used, can be derived from a truncated bivariate normal distribution (TBVN) \cite{arnold1993, kim2006}. The CDF of the SNTN can be found by solving the CDF of two standard bivariate normals, for which there are fast analytical solvers and closed-form approximations. It is therefore computationally trivial to derive exact p-values and confidence intervals for this distribution. 

The rest of the paper is as follows: section \ref{sec:background2} provides the notation for PoSI and data carving,  section \ref{sec:main3} formalizes the relationship between data carving and the pivotal quantities of the SNTN distribution, section \ref{sec:experiments4} highlights the improvements in power that data carving can obtain on real and simulated data, section \ref{sec:conclusion5} provides concluding thoughts, section \ref{sec:broader6} speaks to this paper's broader impact, and section \ref{sec:appendix} contains the appendix.

\section{Background \label{sec:background2}}
\subsection{Post-selection inference (PoSI) \label{sec:posi}}
Consider the standard linear modelling task where 

\begin{align}
y_i &= g(x_i) + e_i, \hspace{3mm} e_i \sim N(0, \sigma^2) \label{eq:dgp}
\end{align}

the data generating process (DGP) in \eqref{eq:dgp} is to be modelled using $n$ IID rows of $x_i \in \mathbb{R}^p$. The least squares estimator is used to perform inference on $g$: $(X^TX)^{-1}X^Tg$. When $g$ is a linear combination of $x_i$: $g_i=\sum_{j=1}^p x_{ij}\beta_j^0$, then the statistic $\hat\beta = (X^TX)^{-1}X^Ty|X \sim N(\beta^0, \sigma^2 (X^TX)^{-1})|X$ has an exact distribution.\footnote{This paper will assume that $\sigma^2$ is known, unless stated otherwise.} 

The statistical quantity of interest is a one-dimensional term $\eta_j^T g=\beta^0_j$, which is the inner product of a row from the matrix $[(X^TX)^{-1}X^T]_{j:}$ and the signal component of the response.  When $j$ is specified independently of the data, then classical pivots can be created, 

\begin{align}
\Phi\Bigg(\frac{\hat\beta_j - \beta_j^0}{\sqrt{\sigma^2 (X^TX)^{-1}_{jj}}} \Bigg) &\sim U(0,1), \label{eq:pval} \\
P( \beta_j^0 \in [\hat\beta_j \pm \sigma (X^TX)^{-1}_{jj} \Phi^{-1}_{1-\alpha/2}] ) &= 1-\alpha  \label{eq:ci}
\end{align}

which obtains exact p-values \eqref{eq:pval} and confidence interval coverage \eqref{eq:ci} at some pre-specified type-I error rate $\alpha$. The functions $\Phi$ and $\Phi^{-1}$ denote the CDF and quantile functions of the standard normal distribution. 

If $\eta_j$ is a function of $y$ however, then the classical results no longer hold. For example, if a sparsity-inducing algorithm like the Lasso is used to select a subset of features $M = \{j: \beta^{\text{lasso}} \neq 0\} \subset \{1, \dots, p\}$, so that $\eta_M=X_M(X_M^TX_M)^{-1}$ then $\eta^T_{M_j}y=\hat\beta_j$ will no longer have the expected distribution from \eqref{eq:pval} since the response vector was used to inform which of the $j \in \{1, \dots, p\}$ directions to test. If classical z-tests are naively used, then the type-I errors will be inflated since these directions will have more correlation with $y$ than the directions  which were not selected (see examples in the appendix).

For a wide class of algorithms -- including the Lasso, marginal screening, forward stepwise regression, least angle regression, and orthogonal matching pursuit \cite{lee2013, lee2014, tibs2016} -- the selection event $\hat M = M(y)$ can be shown to be equivalent to a partition of the $y \in \mathbb{R}^n$ space.

\begin{align}
    \eta^T_{M_j} y | \{\hat{M} = M \} \hspace{2mm} &\longleftrightarrow \hspace{2mm} \eta^T_{M_j} y | \{ A y \leq b \} \label{eq:bijection} \\
    \eta^T_{M_j} g &= \beta_j^M \nonumber
\end{align}

The bijection of \eqref{eq:bijection} between the selection event and the polyhedral constraints ($A$ and $b$) will be specific to each algorithm. Notice the slight shift of notation from $\beta_j^0$ to $\beta_j^M$ which refers to the fact that the oracle parameter may differ for different submodels $M$. This can occur for reasons of omitted variable bias, and highlights that PoSI only considers inference for a given model. 

\begin{align}
    \hat\beta^{\text{PoSI}}_j &= \eta^T_{M_j}y \sim \text{TN}(\beta_j^{M}, \sigma^2_M \|\eta^T_{M_j}\|^2_2, V^{-}(y), V^{+}(y)) \label{eq:bhat_posi}
\end{align}

For a full description and set of proofs as to how polyhedral conditioning leads to a truncated normal distribution in \eqref{eq:bhat_posi}, as well as definitions for $ V^{-}(y)$ and $ V^{+}(y)$ see \cite{lee2013}. Note again that $\sigma^2_M$ may differ to the variance of the original DGP since $M$ may not contain all the covariates with true non-zero coefficients and thus the noise level could rise ($\sigma^2_M \geq \sigma^2$).

\begin{align}
    F_{\beta_j^M, \sigma_M^2 \|\eta^T_{M_j}\|^2_2}^{V^{-}(y), V^{+}(y)}(\hat\beta_{M_j}) &\sim U(0,1) \label{eq:pval_tnorm} \\
    P(\beta_j^M \in [\hat\beta_{M_j}^{-}, \hat\beta_{M_j}^{+}]) &= 1-\alpha \label{eq:ci_tnorm}\\
    \sup_\mu \hspace{1mm} \{ \mu: F_{\mu, \sigma_M^2 \|\eta^T_{M_j}\|^2_2}^{V^{-}(y), V^{+}(y)}(\hat\beta_{M_j}) \geq 1-\alpha/2 \} &= \hat\beta_{M_j}^{-} \label{eq:root_lb_tnorm} \\
    \inf_\mu \hspace{1mm} \{ \mu: F_{\mu, \sigma_M^2 \|\eta^T_{M_j}\|^2_2}^{V^{-}(y), V^{+}(y)}(\hat\beta_{M_j}) \leq \alpha/2 = 0  \} &= \hat\beta_{M_j}^{+} \label{eq:root_ub_tnorm}
\end{align}

Classical pivots can be created for the truncated normal distribution to create p-values that observe a uniform distribution \eqref{eq:pval_tnorm} and confidence intervals with exact $100(1-\alpha)\%$ coverage \eqref{eq:ci_tnorm}. Confidence intervals for the truncated normal can be found by solving the roots seen in equations \eqref{eq:root_lb_tnorm} and \eqref{eq:root_ub_tnorm}, which have a unique solution because $F_{\mu,\sigma^2}^{a,b}(x)$ is monotonically decreasing with respect to $\mu$  (see Lemma A.1 from \cite{lee2013} for a simple proof).

\subsection{Sample splitting}
Now suppose that the $n$ observations of data ($\{(y_1,x_1),\dots,(y_n,x_n)\}$ are split into two groups, of sizes $n_A + n_B = n$. Suppose that group A is used for hypothesis selection (the ``screening'' set), and group B is used to carry out inference (the ``inference'' set). Statistical tests based on this data splitting approach will return us to the classical z-tests that could be formed from equations \eqref{eq:pval}, \eqref{eq:ci},

\begin{align}
     \hat\beta^{\text{split}}_j &= \eta_{M_A}^T y_B = [(X_{B,M_A}^TX_{B,M_A})^{-1} X_{B,M_A}^T]_{j:} y_B \label{eq:bhat_split} \\
    &\sim N\big(\beta^M_j, \sigma^2_{M_A} (X_{B,M_A}^TX_{B,M_A})^{-1}_{jj} \big), \nonumber
\end{align}

Since $y_B, X_B\indep M(y_A, X_A)$. As $n_B$ increases at the expense of $n_A$, the variance of the unbiased tests that come from \eqref{eq:bhat_split} will decrease, providing more power to reject the null hypotheses of the $|M_A|$ directions. However, the ability to effectively screen true signals will decrease as $n_A$ goes down (see \cite{candes2017} for a discussion of the inevitability of noise covariates being selected by the Lasso).

\subsection{Data carving}
The estimator from \eqref{eq:bhat_split} is statistically wasteful, since it ignores the leftover fisher information found in $y_A | M_A$ \cite{fithian2017}, which as section \ref{sec:posi} showed could still be used to carry out exact statistical tests. 

\begin{align}
    \hat\beta^{\text{carve}}_j &= w_A \hat\beta^{\text{split}}_j + w_B \hat\beta^{\text{PoSI}}_j\label{eq:bhat_carve} \\
    w_j &= n_j / n, \hspace{2mm} j \in \{A,B\} \nonumber
\end{align}

The data carving estimator proposed in this paper is a sample-sized weighted combination of the PoSI and sample splitting estimators. When the distribution of \eqref{eq:bhat_carve} is known under the null, then this estimator will strictly dominate sample splitting in terms of power \eqref{eq:bhat_split} (see Theorem 9 from \cite{fithian2017}).

\section{Main result: data carving with the polyhedral lemma \label{sec:main3}}
The sum of a normal and a truncated normal has a well-defined parametric distribution. Lemma \eqref{lemma:sntn} is a simple restatement without proof of Lemma 1 from \cite{kim2006} with some changes of notation.

\begin{lemma}[Characterizing the sum of a normal and truncated normal (SNTN) distribution \label{lemma:sntn}]
    Define $X_1 \sim N(\mu_1, \tau_1^2)$ and $X_2 \sim \text{TN}(\mu_2, \tau_2^2, a, b)$, then $Z = c_1 X_1 + c_2 X_2$, $c_i \in \mathbb{R}$, is said to follow an SNTN distribution denoted as either $\text{SNTN}(\mu_1, \tau_1^2, \mu_2, \tau_2^2, a, b, c_1, c_2)\overset{d}{=}\text{SNTN}(\theta_1, \sigma_1^2, \theta_2, \sigma_2^2, \omega, \delta)$ distribution where $\theta_1=\sum_{i=1}^2c_i\mu_i$, $\sigma_1^2=\sum_{i=1}^2c_i^2\tau_i^2$, $\theta_2=\mu_2$, $\sigma_2^2=\tau_2^2$, $\rho=c_2\sigma_2/\sigma_1$, $\lambda=\rho/\sqrt{1-\rho^2}$, $\gamma=\lambda/\rho$, $m_j(x)=(x-\theta_x)/\sigma_x$, $\omega=m_2(a)=(a-\theta_2)/\sigma_2$, $\delta=m_2(b)=(b-\theta_2)/\sigma_2$. The PDF ($f$) and CDF ($F$) of the SNTN distribution can be characterized as follows:

    \begin{align}
        f_{\theta,\sigma^2}^{\omega,\delta}(z) &= \frac{\phi(m_1(z))[\Phi(\gamma\delta -\lambda m_1(z)) - \Phi(\gamma\omega -\lambda m_1(z))]}{\sigma_1[\Phi(\delta)-\Phi(\omega)]} \label{eq:sntn_pdf} \\
        F_{\theta,\sigma^2}^{\omega,\delta}(z) &= \frac{B_\rho(m_1(z),\delta) - B_\rho(m_1(z),\omega)}{\Phi(\delta)-\Phi(\omega)}  \label{eq:sntn_cdf}
    \end{align}

    Where $B_\rho(x_1, x_2)$ is the CDF of a standard bivariate normal with correlation $\rho$.
\end{lemma}

The CDF of the SNTN is written in terms of the bivariate normal CDF rather than its orthant probabilities as has historically been done \cite{kim2006}. The appendix provides details on fast and efficient methods for calculating \eqref{eq:sntn_cdf}.

\begin{lemma}[An exact distribution of data carving estimator \label{lemma:carving}]
    Let $y \sim N(X\beta^0, \tau^2 I_n)$ where $y \in \mathbb{R}^n$ and $x_i \in \mathbb{R}^p$. Consider any selection algorithm which generates a subset of hypotheses $M(y) \subset \{1,\dots,p\}$ which is in bijection to an affine set of constraints on $y$: $\{\hat{M} = M(y)\} \longleftrightarrow \cup_s \{  A_s y \leq b_s \}$. Suppose the data is split into two non-overlapping sets: $n_A \subset \{1,\dots, n\}$ and $n_B \subset \{1,\dots, n\} \setminus n_A$, where neither set is empty $n_A \neq \emptyset \land n_B \neq \emptyset$. Suppose set A is used to generate $M_A=\hat{M}=M(y_A)$, and both sets of data are used to perform linear regression on these $|\hat{M}|$ directions. Then the weighted average of the linear regression contrasts follows an SNTN distribution.

    \begin{align*}
        \hat\beta_j &= n^{-1} [n_A \eta_{A,M_j}^Ty_A + n_B \eta_{B,M_j}^Ty_B] \\
        \eta_{M_j}^T &= [(X_{M_A}^TX_{M_A})^TX_{M_A}^T]_{j:} \\
        \hat\beta_j &\sim \text{SNTN}(\theta_1, \sigma_1^2, \theta_2, \sigma_2^2, \omega, \delta) \\
        c_1 &= n_B/n, \hspace{2mm} c_2=n_A/n \\
        \theta_1 &= \theta_2 = \beta_j^M \\
        \sigma_1^2 &= \frac{\tau_M^2}{n^2} \big[n_B^2(X_{B,M_A}^TX_{B,M_A})_{jj}^{-1} + n_A^2(X_{A,M_A}^TX_{A,M_A})_{jj}^{-1} \big]    \\
        \sigma_2^2 &= \tau^2_M (X_{A,M_A}^TX_{A,M_A})_{jj}^{-1} \\
        \omega &= [V^{-}(y_A)-\beta_j^M]/ [\tau_M(X_{A,M_A}^TX_{A,M_A})_{jj}^{-1/2}] \\
        \delta &= [V^{+}(y_A) -\beta_j^M]/ [\tau_M(X_{A,M_A}^TX_{A,M_A})_{jj}^{-1/2}] \\
        \rho &= \frac{\sqrt{n \cdot n_A} \cdot 
        (X_{A,M_A}^TX_{A,M_A})^{-1/2}}{\big[n_B^2(X_{B,M_A}^TX_{B,M_A})^{-1} + n_A^2(X_{A,M_A}^TX_{A,M_A})^{-1} \big]^{1/2}}
    \end{align*}

    Where the terms above follow from the fact that PoSI coefficient  $\eta_{A,M_j}^Ty_A \sim \text{TN}(\beta_j^M,\tau^2_M (X_{A,M_A}^TX_{A,M_A})^{-1}_{jj}, V^{-}(y_A), V^{+}(y_A))$ follows a truncated normal because of the polyhedral lemma shown in \cite{lee2013}, and the inference-set coefficient $\eta_{B,M_j}^Ty_B \sim N(\beta_j^M, \tau^2_M (X_{B,M_A}^TX_{B,M_A})^{-1}_{jj})$ is normally distribution due to the independence generated by sampling splitting. 
\end{lemma}

Lemma \eqref{lemma:carving} puts the pieces together for combining the truncated normal distribution of \eqref{eq:bhat_posi} and the Gaussian distribution of \eqref{eq:bhat_split} for a data carving estimator of $\beta_j^M$ as seen in \eqref{eq:bhat_carve}. 

\begin{theorem}[Exact p-values and confidence intervals for data carving \label{thm:exact}]
    Using the same notation and assumptions from Lemma \eqref{lemma:carving}, define $\beta_j^M = \eta^T_{M_j} g$, where $g$ is from \eqref{eq:dgp}. Then it follows that,
    
    \begin{align}
        F_{\beta_j^M, \sigma^2}^{\omega, \delta}(\hat\beta_j) &\sim  U(0,1) \label{eq:pval_carve}
    \end{align}

    The data carving estimator $\hat\beta_j$ has uniform p-values under the null of $\theta_1=\theta_2=\beta_j^M$,

    \begin{align}
        P(\beta_j^M \in [\hat\beta_j^{-}, \hat\beta_j^{+}]) &= 1-\alpha \label{eq:ci_sntn} \\
        \hat\beta_j^{-} &= \sup_\mu \hspace{1mm} \{ \beta_j: F_{\beta_j, \sigma^2}^{\omega, \delta}(\hat\beta_j) \geq 1-\alpha/2 \} \nonumber  \\
        \hat\beta_j^{-} &= \inf_\mu \hspace{1mm} \{ \beta_j: F_{\beta_j, \sigma^2}^{\omega, \delta}(\hat\beta_j) \leq \alpha/2 = 0  \} \nonumber
    \end{align}

    And that exact confidence intervals can be found by solving for the roots for the CDF. Since $F_{\beta_j, \sigma^2}^{\omega, \delta}(z)$ is monotonically decreasing with respect to $\beta_j$ (see the appendix for a proof), it holds that the solutions to the roots are unique.

\end{theorem}

Theorem \eqref{thm:exact} shows that the usual pivotal quantities used in frequentist inference can be created for the data carving estimator \eqref{eq:bhat_carve}. As a reminder, $F_{\beta, \sigma^2}^{\omega, \delta}(z)$ for the SNTN can be found by solving two bivariate normal CDFs making the calculation of exact p-values computationally trivial. The calculation of the confidence intervals can be done by repeatedly evaluating the CDF until an exact solution is found, which can be easily done using off-the-shelf root-finding algorithms (see the appendix for more details). 

\section{Experiments \label{sec:experiments4}}
\subsection{Sample mean \label{sec:xbar}}
Before looking at multivariate models, consider the case of estimating a sample mean in the presence of the ``file-drawer problem.'' Suppose there are $n$ observations: $y \sim \text{MVN}(\mu, \sigma^2 I_p)$, and only coordinates which have an average value of at least one will be tested: $M = \{j: n^{-1}\sum_{i}^n y_{ij} > 1\}$. In this case $X=[1]^n$ is an intercept, and up to $p$ directions of $y$ will be tested. A one-sided null hypothesis of $H_0: \mu_j \leq 0$ will be used.

It is easy to see that for the conditioning event $\{\hat{M} = M\} = \{A y \leq b\}$, $A$ is a $(p,n)$ matrix of values $-1/n$, and $b$ is a vector of negative ones. 

\begin{align*}
    \bar{y}_j &= w_A \cdot y_{Aj} | \bar{y}_{Aj} > 1 + w_B\cdot\bar{y}_{Bj} \\
    &\sim \text{SNTN}(\mu_j, \sigma^2_B,  \mu_j, \sigma^2_A, 1, \infty, w_B, w_A) \\
    \sigma^2_k &= \sigma^2 / n_k, \hspace{2mm} k \in \{A,B\}
\end{align*}

For a given choice of $n$, $n_A$, and $\sigma^2$, a critical value under the null hypothesis can be determined: $c_\alpha = \inf_x F_{0,\sigma^2}^{m_2(1), \infty}(x) - (1-\alpha) \geq 0$, and the power can be calculated across a range of $\theta_1=\theta_2=\mu$ values: $1-\beta = 1-F_{\mu,\sigma^2}^{m_2(1), \infty}(c_\alpha)=P(\bar{y}_j > c_\alpha)$, where $\beta$ is the type-II error rate, since the null hypothesis is rejected whenever $\bar{y}_j > c_\alpha$.

\begin{figure}[!htb]
    \centering
    \includegraphics[width=0.8\linewidth]{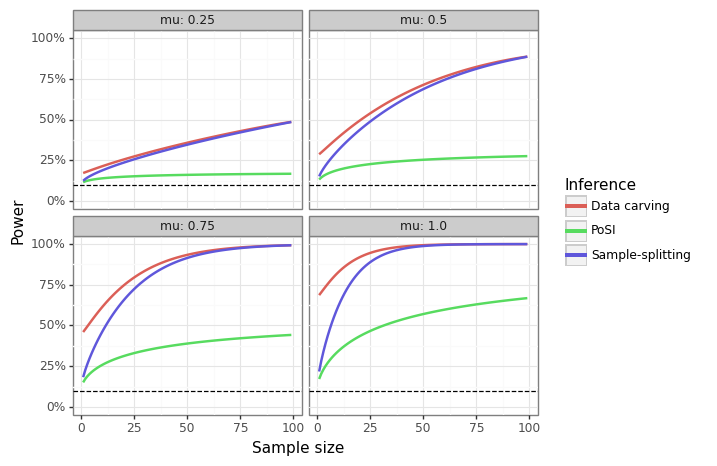}
    \caption{Data carving for sample mean estimation. The horizontal axis ranges from $n_{\{A,B\}} \in \{1,\dots,99\}$, where $\sigma^2=4$, $n=100$, and $\alpha=0.1$. Power is calculated using the $1-\alpha$ quantile (for a one-sided hypothesis test) of the null distribution ($H_0:\mu\leq 0)$ and evaluating the CDF of the alternative distribution for different values of $n_{\{A,B\}}$, and $\mu$. \label{fig:sample_mean}}
\end{figure}

Figure \ref{fig:sample_mean} shows the improvement in power that data carving has over both sample splitting and PoSI. When the signal size is small, sample-splitting and data carving dominate PoSI, and carving is only marginally better than sample splitting. However, as $\mu$ increases, data carving is shown to be effective with only a small amount of data dedicated to the inference set. 

\subsection{High dimensional sparse regression \label{sec:hdi}}
Consider the high dimensional linear regime of \eqref{eq:dgp} where $p > n$, but $S=\{j: \beta_j\neq 0\}$, $s=|S| < n$. Define the signal-to-noise ratio (SNR) as: $\text{var}(X \beta^0) / \sigma^2$. In the simulation experiments below, $n=100$, $p=150$, $s=5$, $\sigma^2=1$, and $\alpha=0.1$. For different levels of the SNR, the first 5 coefficients of $X$ will have a fixed, positive, and non-zero value. 

Two different sparsity inducing algorithms were considered for each simulation draw: the Lasso and marginal screening. For the lasso, $\lambda$ is set to be 72.5\% of $\lambda_{\text{max}} = \inf_\lambda |\{j: \beta_j^{\text{lasso}} \neq 0 \}| = 0 $. For marginal screening, the top $k=10$ features are selected (i.e. the features with the highest linear correlation with $y$). 

\begin{figure}
     \centering
     \begin{subfigure}[b]{0.65\textwidth}
         \centering
         \includegraphics[width=\textwidth]{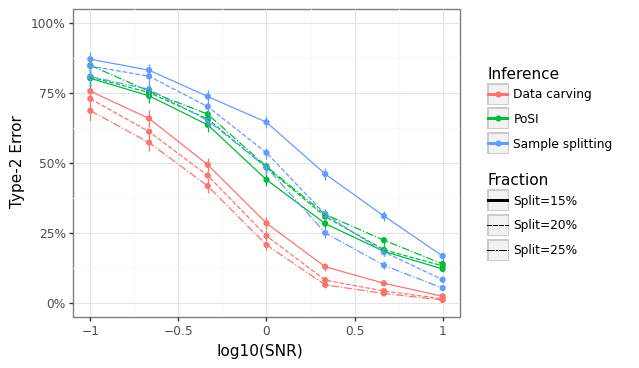}
         \caption{Lasso (Type-II error)}
         \label{fig:typeII_lasso}
     \end{subfigure}
     \vskip\baselineskip
     \begin{subfigure}[b]{0.65\textwidth}
         \centering
         \includegraphics[width=\textwidth]{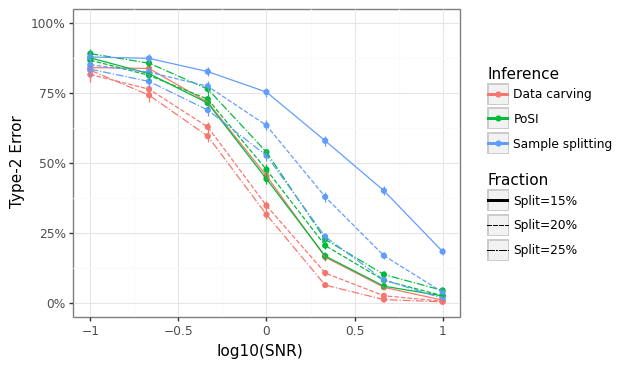}
         \caption{Marginal screening (Type-II error)}
         \label{fig:typeII_screening}
     \end{subfigure}
    \caption{Type-II error (1-power) for data carving against other approaches. Line ranges show 95\% confidence intervals for a binomial proportion using the Clopper–Pearson method based on 500 simulation runs ($n=100$, $p=150$, $s=5$, $\sigma^2=1$, and $\alpha=0.1$)}
    \label{fig:hdi_power}
\end{figure}

Experiments were carried out for 7 different SNRs on the log10 scale: $\{-1,-2/3,-1/3,0,1/3,2/3,1\}$, and 3 different fractions used for the sample splitting: $n_B/n = \{0.15, 0.20, 0.25\}$. Each of the 21 combinations had 500 simulations runs. Figure \ref{fig:hdi_power} shows that there were significant gains in power using PoSI, especially for moderate levels of the SNR (over 25pp). 

As was seen in the case of the sample mean estimation, a small amount of inference data helps to improve power by a substantial amount, while further contributions to $n_B$ have diminishing effects. For example, Figure \ref{fig:typeII_lasso} shows that for the Lasso when the log10(SNR)=0, using 15\% of the data for carving yields a power improvement of  16pp and 37pp over PosI and sample splitting, whilst providing another 10\% of data for carving increases its absolute power by only 5pp. The supplementary section provides further results in terms of screening probabilities and type-I error rates (which are shown to be controlled and at the expected level).

\subsection{Diabetes dataset \label{sec:diabetes}}
\begin{figure}[!htb]
    \centering
    \includegraphics[width=0.7\linewidth]{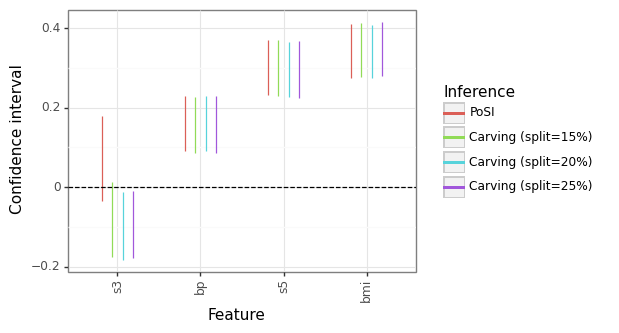}
    \caption{Data carving vs PoSI using the Lasso on the diabetes dataset ($n=442$ and $p=10$). Confidence intervals use $\alpha=0.1$ and $\sigma^2$ was estimated using 5-fold cross-validation. \label{fig:diabetes}}
\end{figure}

For a real-world data example, a Lasso algorithm with $\lambda_{\text{max}}=0.25$ was used on the \texttt{diabetes} dataset \cite{efron2004}. A total of five features were selected by the full-sample Lasso, as well as data-carving models for $n_B/n = \{0.15, 0.20, 0.25\}$. Figure \ref{fig:diabetes} shows the confidence intervals that were generated at the 90\% level. The data carving procedures found one additional feature to be statistically significant -- \texttt{s3} features (high-density lipoproteins) -- compared to the PoSI approach. As the simulation data showed, in the regime of moderate signal size, we should expect data carving to discover more true positives.

\section{Conclusion \& limitations \label{sec:conclusion5}}
This paper has shown that test statistics produced by the data carving procedure follow a parametric distribution when the selection event can be represented by an affine set of constraints on a Gaussian response vector. This parametric distribution, the sum of a normal and a truncated normal (SNTN), has a CDF that can be found by  calculating two bivariate normal CDFs, which can be done trivially with modern statistical software. Simulation and real-world data show that data carving can produce substantial improvements in power (a fact which echoes theoretical and empirical evidence found by other papers). Data carving can be applied to popular algorithms like the Lasso, marginal screening, or forward stepwise regression. 

Packages like \href{https://cran.r-project.org/web/packages/selectiveInference/index.html}{\texttt{selectiveInference}} have helped increase access of PoSI methods for data scientists. This work seeks to augment existing tools, and the \href{https://pypi.org/project/sntn/}{\texttt{sntn}} package provides examples of how to use the SNTN distribution to carry out exact inferences for data carving. 

There are several limitations to this paper. First, it does not address the ``p-value lottery'' problem that comes from randomly splitting the data. Different data splits can therefore yield different inferences. Since the SNTN distribution easily generates exact p-values under the null hypothesis, random variations introduced by splitting could be ameliorated by the multicarving framework suggested by \cite{buhlmann2021}. Second, in order to obtain an exact distribution, this paper assumed that the response was Gaussian and that the variance was known. In reality, estimating $\sigma^2$ can be challenging, especially in high-dimensional settings (see \cite{reid2016}). However, this nuisance parameter challenge is present in traditional PoSI approaches as well. Lastly, this paper does not consider generalized linear models (GLMs) which are used to model binary, count, or survival time outcomes. A natural extension would be to use the normal asymptotics which has been proposed for PoSI-GLMs \cite{taylor2016b} as an approximation for the SNTN.

\section{Broader impacts \label{sec:broader6}}
The method developed in this paper provides a computationally tractable way to improve the power of statistical tests for certain adaptive hypothesis testing procedures. There are unlikely to be risks of such a tool to the broader community. However, like all statistical tools, this one is based on numerous assumptions. Data carving with the SNTN will only be ``exact'' when the data follows the Gaussian distribution described. Furthermore, modelling any phenomena with linear regression will naturally incorporate any biases or confounding relationships which are found in the data. The approach outlined in this paper does not address any of the traditional concerns and pitfalls that go along with regression modelling. 

Thinking carefully about how hypotheses get generated and whether they represent novel findings can also be understood in the broader context of the reproducibility (or replication) crisis \cite{navarro2020}. Data carving can help reduce false positives by forcing practitioners to think carefully about how they construct their hypothesis-generating pipeline. However, this method still leaves ``researcher degrees of freedom'' on the table \cite{simmons2011}. Data carving and PoSI do not change the need to follow the best practices of open science including pre-registration and other forms of hypothesis testing transparency. 

\medskip
\bibliography{ref}
\newpage

\section{Appendix \label{sec:appendix}}
The appendix is structured as follows: section \ref{sec:reproducibility} links to the code needed to reproduce the analysis found in this paper,  section \ref{sec:cdf_integral} discusses the computational considerations for calculating the CDF of the SNTN distribution along with other root finding algorithms, section \ref{sec:appendix_figs} provides additional figures from the simulation results, and section \ref{sec:proofs} gives proofs related to the main theorem of the paper.

\subsection{Reproducibility \label{sec:reproducibility}}

Methods from the \href{https://pypi.org/project/sntn/}{\texttt{sntn}} package, co-released with this paper, were used to generate all of the main results. The \href{https://github.com/ErikinBC/sntn/tree/main/simulations}{simulations} folder from the main github page contains the scripts needed to generate all the figures seen in this paper:

\begin{enumerate}
    \item \href{https://github.com/ErikinBC/sntn/blob/main/simulations/2_sample_mean.py}{Sample mean} results for section \ref{sec:xbar}
    \item \href{https://github.com/ErikinBC/sntn/blob/main/simulations/3_marginal_screening.py}{Marginal screening} and \href{https://github.com/ErikinBC/sntn/blob/main/simulations/4_lasso.py}{Lasso} PoSI results for section \ref{sec:hdi}
    \item \href{https://github.com/ErikinBC/sntn/blob/main/simulations/5_diabetes.py}{Diabetes} results for the real-word dataset in section \ref{sec:diabetes}
    \item \href{https://github.com/ErikinBC/sntn/blob/main/simulations/0a_sim_bvn.py}{BVN-CDF} runtime and accuracy experiments see in section \ref{sec:bvn_cdf}
    \item \href{https://github.com/ErikinBC/sntn/blob/main/simulations/0b_sim_tnorm.py}{Root-finding} runtime and accuracy experiments see in section \ref{sec:rootfinding}
\end{enumerate}

\subsection{Integral computations \label{sec:cdf_integral}}
Performing inference on the SNTN distribution requires three computational procedures:

\begin{itemize}
    \item A closed-form calculation for the PDF, $f_{\theta,\sigma^2}^{\omega,\delta}(z)$:  \eqref{eq:sntn_pdf}
    \item Two bivariate normal calculations for the CDF, $F_{\theta,\sigma^2}^{\omega,\delta}(z)$: \eqref{eq:sntn_cdf} 
    \item Root finding for the confidence intervals and quantiles, $\hat\beta_j^{\{-,+\}}$: \eqref{eq:ci_sntn}
\end{itemize}

The PDF of the SNTN distribution can be trivially calculated using the built-in PDF and CDF functions of the normal distributions found in every statistical software (e.g. \texttt{scipy.stats.norm.\{pdf,cdf\}}), and is therefore not discussed here. 

\subsubsection{Bivariate normal CDF \label{sec:bvn_cdf}}

Several approaches were considered for obtaining estimates of the CDF of the bivariate normal distribution:

\begin{itemize}
    \item Monte Carlo estimation \cite{genz1992} using by \texttt{scipy.stats.multivariate\_normal} (denoted ``scipy'')
    \item Owen's-T Function \cite{owen1956} using \texttt{scipy.special.owens\_t} (denoted ``owen'')
    \item Cox's approximation using equation (3) from \cite{cox1991} (denoted ``cox1'')
    \item Monte Carlo estimation of the conditional probability $P(X_1\geq x_1, X_2\geq x_2)=\Phi(-x_1)E[\Phi((\rho X_1 - x_2)/\sqrt{1-\rho^2} | X_1 > x_1]$, by drawing from $X_1 \sim \text{TN}(0,1,x_1,\infty)$ (denoted ``cox2'')
    \item Gaussian quadrature using \texttt{scipy.integrate.fixed\_quad} for two different transformations proposed by \cite{drezner1990}: $\Phi(-x_1)\Phi(-x_2) + \frac{1}{2\pi} \int_0^{\arcsin(\rho)} \exp[ -(x_1^2 + x_2^2 - 2x_1x_2\sin(\theta))/(2 \cos^2(\theta))] d\theta$ (denoted ``drezner1'') and $\Phi(-x_1)\Phi(-x_2) + \frac{1}{2\pi} \int_0^\rho \frac{1}{1-r^2} \exp[ (-x_1^2 + k^2 - 2x_1x_2r)/(2 (1-r)^2)] dr$ (denoted ``drezner2'')
\end{itemize}

\begin{figure}[!htp]
    \centering
    \includegraphics[width=0.75\textwidth]{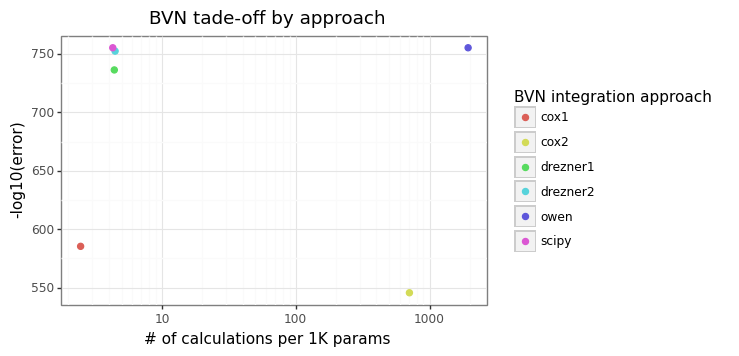}
    \caption{Run-time vs accuracy for the bivariate normal CDF approaches. The horizontal axis is measured in seconds of time, and the vertical axis is the final cumulative error. The ``cox2'' approach used 1000 simulation draws. Error is based of off 175 randomly chosen samples, whereas runtime based on total time for each method across 54K parameter combination evaluations.}
    \label{fig:bvn_tradeoff}
\end{figure}

Each of the six approaches listed above was tested for 54000 different bivariate normal CDF combinations, where each simulation had a unique combination of the 5 bivariate normal parameters: $\text{B}(\mu_1,\mu_2,\sigma_1^2,\sigma_2^2,\rho)$. One random data point ($x_1,x_2)$ was drawn from each of these distributions, and a ground-truth CDF was determined by sampling 10000000 points from the distribution and calculating the empirical CDF. Figure \ref{fig:bvn_tradeoff} shows how each of the approaches fared in terms of both run-time and accuracy. The ``owen'' and ``scipy'' approaches were the most accurate, but the former was up to 1000x faster. 

Overall, Owen's-T method is used as the default solver in the \texttt{sntn} package for calculating the CDF of the SNTN distribution, which the slower but more reliable ``scipy'' method being used a a backup. 

\subsubsection{Root-finding algorithms \label{sec:rootfinding}}

Finding confidence intervals or quantiles of the SNTN distribution, amounts to solving a root. For the case of the SNTN confidence interval \eqref{eq:ci_sntn}, this amount to finding a value of $\mu_1=\mu_2=\mu$ such that $\mu: F_{\mu,:}^{:,:}(z) - \alpha = 0$, holding the other parameters fixed for some point $z$. Alternatively, the quantile function seeks to find a value $z$ such that holding all parameters constant yields a CDF matching some percentile: $z: F_{:,:}^{:,:}(z) - p = 0$.

Experiments were run on the slightly more challenging root-finding problem of the truncated normal confidence interval, since the upper/lower bounds can be harder to solve as $\partial F / \partial \mu \approx 0$ for values of $\mu \ll a$ or $\mu \gg b$. In other words, if a root solver was efficient for the truncated normal, it was expected to be so for the SNTN distribution as well. 

\begin{figure}[!htp]
    \centering
    \includegraphics[width=0.75\textwidth]{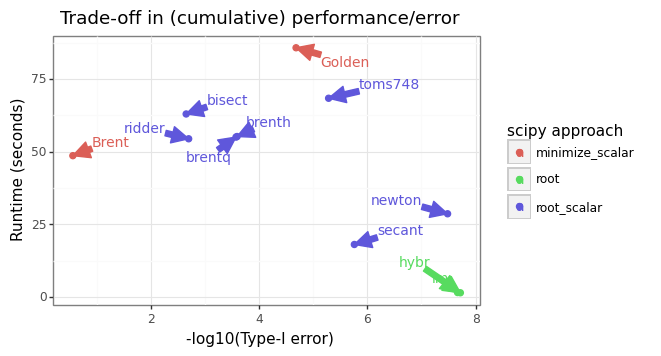}
    \caption{Run-time vs accuracy for root-finding approaches. Error is cumulative based on evaluating $\sum_i |F_{\mu,:}^{:,}(z) - \{\alpha/2,1-\alpha/2\}|$. Roots are based on the confidence intervals needed for a truncated normal distribution.}
    \label{fig:root_tradeoff}
\end{figure}

Figure \ref{fig:root_tradeoff} shows the trade-off between the accuracy and runtime for solving 1050 two-sided confidence intervals at the 5\% level (i.e. 2100 roots). Overall the vectorized implementations of the \texttt{scipy.optimize.root} with a method of ``hybr'' out-performed all other approaches. For this reason, the confidence interval and quantile methods of the \texttt{sntn} package us the ``root'' approach as default, and a bounded bisection method as backup if the root cannot be found in the first try.

\subsection{Figures \label{sec:appendix_figs}}
\begin{figure}[!htp]
     \centering
     \begin{subfigure}[b]{0.75\textwidth}
         \centering
         \includegraphics[width=\textwidth]{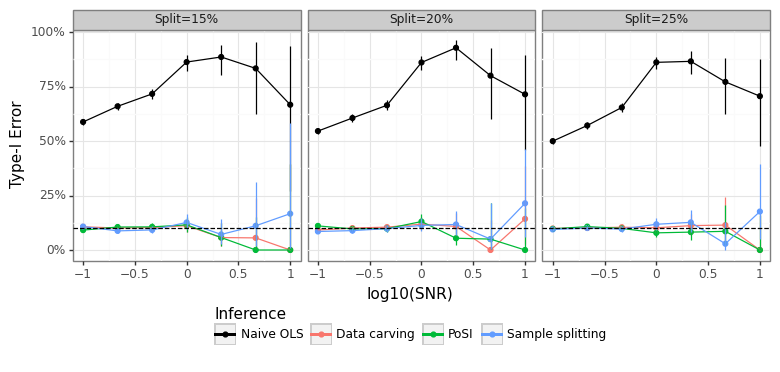}
         \caption{Lasso (Type-I error)}
         \label{fig:typeI_lasso}
     \end{subfigure}
     \vskip\baselineskip
     \begin{subfigure}[b]{0.75\textwidth}
         \centering
         \includegraphics[width=\textwidth]{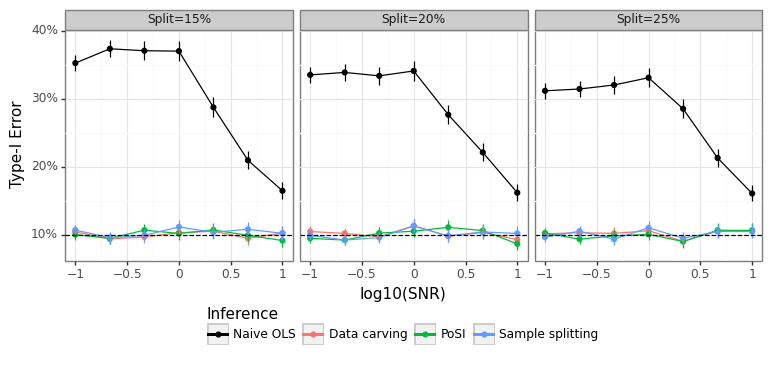}
         \caption{Marginal screening (Type-I error)}
         \label{fig:typeI_screening}
     \end{subfigure}
    \caption{Type-I error (1-power) for PoSI approaches. Line-ranges show 95\% CIs for a binomial proportion using the Clopper–Pearson method based on 500 simulation runs ($n=100$, $p=150$, $s=5$, $\sigma^2=1$, and $\alpha=0.1$). ``Naive OLS'' refers to classical p-values generated from running an OLS on the screening data that was used to fit the lasso or marginal screening model.}
    \label{fig:type1}
\end{figure}

This section contains additional figures related to the simulations discussed in section \ref{sec:experiments4}. Figure \ref{fig:type1} shows the Type-I error rate for the different PoSI approaches. The Naive OLS approach, which is based on a z-test from running an OLS on the same data from which the covariates were selected by the lasso or marginal screening algorithm, shows type-I errors well above the expected nominal level. This is be expected since these features were selected because they had correlations higher than the others. In contrast, the data carving, PoSI (i.e. 100\% screening), and sample splitting methods all show type-I error rates around the expected level (10\%). The confidence intervals around the binomial proportion (the type-I error) are large when the SNR is large for the lasso because few false positives are selected meaning that a small share of the 500 simulation runs have any false positives. 

\begin{figure}[!htp]
     \centering
     \begin{subfigure}[b]{0.45\textwidth}
         \centering
         \includegraphics[width=\textwidth]{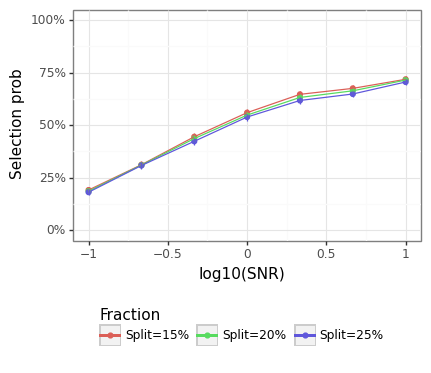}
         \caption{Lasso}
         \label{fig:selprob_lasso}
     \end{subfigure}
     \begin{subfigure}[b]{0.45\textwidth}
         \centering
         \includegraphics[width=\textwidth]{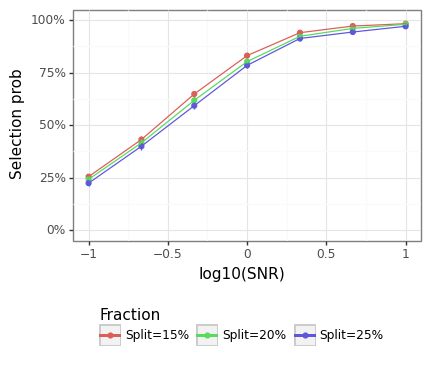}
         \caption{Marginal screening}
         \label{fig:selprob_screening}
     \end{subfigure}
    \caption{Precision for PoSI approaches. Selection prob refers to the probability that a selected covariate is a true positive ($n=100$, $p=150$, $s=5$, $\sigma^2=1$, and $\alpha=0.1$).}
    \label{fig:selprob}
\end{figure}

Figure \ref{fig:selprob} shows the change in the ability of the lasso and marginal screening algorithms to screen true positives compared to false positives. Specifically, the figure shows the probability that a selected covariate is a true positive (i.e. the precision). The figure shows that for both algorithms, the increase in screening accuracy is dominated by the underlying signal size, rather than by the share of data which goes towards the screening dataset. This also highlights that when the share of data given to the inference set goes up, there will be many instances where the change in power more than offsets the decline in the selection power. 

\subsection{Proofs \label{sec:proofs}}
The main results of this paper come from Theorem \ref{thm:exact}, which says that we can expect uniform p-values when the SNTN estimator is evaluated under the null hypothesis, and that the confidence intervals will be exact. 

\subsubsection{Uniform p-values}

The proof to \eqref{eq:pval_carve} is trivial since it relies on the well-known fact that when the CDF of a test statistic is evaluated under the null hypothesis it will follow a uniform distribution. A quick restatement of some of the notation from Lemma \ref{lemma:sntn} and \ref{lemma:carving} will be given for convenience.

For two sets of data $n_A$ and $n_B$, two estimators are fit by running OLS on the selected subset of covariates $M_A$, and their weighted average follows an SNTN distribution.

\begin{align*}
    &\text{From the polyhedral lemma \cite{lee2013}} \\
    \hat\beta^A_j &\sim \text{TN}(\beta_j^M,\tau^2_M (X_{A,M_A}^TX_{A,M_A})^{-1}_{jj}, V^{-}(y_A), V^{+}(y_A)) \\
    &\text{From the independence of sample splitting} \\
    \hat\beta^B_j &\sim N(\beta_j^M, \tau^2_M (X_{B,M_A}^TX_{B,M_A})^{-1}_{jj}) \\
    &\text{From the definition of an SNTN - Lemma \ref{lemma:sntn}} \\
    \hat\beta_j &= \frac{n_B}{n}\hat\beta^B_j +  \frac{n_A}{n}\hat\beta^A_j \sim \text{SNTN}(\theta_1, \sigma_1^2, \theta_2, \sigma_2^2, \omega, \delta)
\end{align*}

Consider the following null hypothesis: $H_0: g_j=\beta_j$ where $g=g_j(x_j) + g_{-j}(x_{-j})=\beta_j x_j + g_{-j}(x_{-j})$ from \eqref{eq:dgp}, then it follows that $\hat\beta_j \sim \text{SNTN}(\theta_1, \sigma_1^2, \theta_2, \sigma_2^2, \omega, \delta)$ under the null hypothesis because $\theta_2=\beta_j$, and $\theta_1=(n_B/n)\beta_j + (n_B/n)\beta_j=\beta_j$. Define $F_{\theta,\sigma^2}^{\omega,\delta}$ as the CDF of $\hat\beta_j$, then consider the random variable $P(p)=Pr(F_{\theta,\sigma^2}^{\omega,\delta}(\hat\beta_j) < p)=p\sim U(0,1)$ by definition of the CDF of a uniform distribution. This clearly shows that $P$ is uniform under the null. $\square$

\subsubsection{Confidence intervals}

Proof that the confidence intervals for the SNTN are exact \eqref{eq:ci_sntn}, will require demonstrating that the CDF is a monotonically decreasing function with respect to $\mu$. This will imply an exact solution to the root-finding problem, and coverage properties that will emerge from the fact that only values in less than the ($\alpha/2$)-quantile or greater than the ($1-\alpha/2$)-quantile will not be covered by the confidence interval.

Let $Z \sim F_{\theta,\sigma^2}^{\omega,\delta}$ denote the CDF of an SNTN distribution as defined in Lemma \ref{lemma:sntn}, where $Z=X_1+X_2$. Assume that $X_1\sim N(\mu,\tau_2^1)$ and $X_2\sim \text{TN}(\mu,\tau_2^2,a,b)$ have the same mean, $\mu_1=\mu_2=\mu$ and that $c_1 + c_2=1$. It follows that $\theta=(\theta_1,\theta_2)=(c_1\mu_1+c_2\mu_2,\mu_2)=(\mu,\mu)$. Let $G_\mu=F_{\mu,\tau^2_1}$ and $T_\mu=F_{\mu,\tau^2_1}^{a,b}$ denote the CDFs of the normal and truncated normal, respectively. It is easy to see that,

\begin{align*}
    \frac{\partial G_\mu(x)}{\partial \mu} &= -\frac{1}{\sigma}\frac{1}{\sigma \sqrt{2\pi}} \exp\Big\{-\frac{(x-\mu)^2}{2\sigma^2} \Big\} \\
    &< 0, \hspace{2mm} \forall x \in \mathbb{R}, \mu \in \mathbb{R}, \sigma \in \mathbb{R^+} 
\end{align*}

Showing that the Gaussian CDF is monotonically decreasing w.r.t. $\mu$ across the support of the distribution. Lemma A.1 from \cite{lee2013} gives a proof as to why $\partial T_\mu / \partial \mu < 0$, $\forall x \in \mathbb{R}, \mu \in \mathbb{R}, \sigma \in \mathbb{R^+}, b > a, b \in \mathbb{R}, a \in \mathbb{R}$. It is well known that a smooth continuous function whose derivative is negative at all points is monotonically decreasing, demonstrating that both $G$ and $T$ are strictly decreasing functions.

Since $G_{\mu'}(x) < G_\mu(x)$, $\forall \mu' > \mu$ and $T_{\mu'}(x) < T_\mu(x)$, $\forall \mu' > \mu$, it is easy to see that:

\begin{align*}
    G_{\mu'}(x) + T_{\mu'}(x) &< G_{\mu}(x) + T_{\mu}(x) \\
    \underbrace{[G_{\mu'}(x)-G_{\mu}(x)]}_{<0} + \underbrace{[T_{\mu'}(x)-T_{\mu}(x)]}_{<0} &< 0 \hspace{2mm} \longleftrightarrow \\
    F_{\mu',\sigma^2}^{\omega,\delta} - F_{\mu,\sigma^2}^{\omega,\delta}  &< 0 
\end{align*}

Because $F_{\mu,\sigma^2}^{\omega,\delta}$ is a continuous function between $[0,1]$, and is strictly decreasing w.r.t. to $\mu \in \mathbb{R}$, by the intermediate value theorem,

\begin{align*}
    \mu^*(\alpha, z) &= \inf_\mu \hspace{2mm} \{\mu: F_{\mu,\sigma^2}^{\omega,\delta}(z) \leq \alpha\},
\end{align*}

$\mu^*$ must be unique. Because $G$ and $T$ are monotonically increasing functions w.r.t. $x \in \mathbb{R}$, it also holds that $F$ is monotonically increasing w.r.t. $x$, and that the $\alpha$-quantile function is unique:

\begin{align*}
    z^*(\alpha, \mu) &= \sup_z \hspace{2mm} \{z: F_{\mu,\sigma^2}^{\omega,\delta}(z) \leq \alpha\}.
\end{align*}

Next consider a one-sided confidence interval $(-\infty, z^{+})$ where $z^{+} = \mu^*(1-\alpha, z)$,

\begin{align*}
    P(\mu \in (-\infty, z^{+})) &= P(z^{+} \geq \mu) \\
    &= P(\mu^*(1-\alpha, z) \geq \mu) \\
    &= 1-\alpha
\end{align*}

Since $\mu^*(\alpha, z^*(\alpha, \mu))=\mu$. By a similar logic, $P(\mu \in (z^{-}, \infty)) = 1-\alpha$ where $z^{-} = \mu^*(\alpha, z)$. Thus the two-sided confidence interval will be exact: $P(\mu \in (z^{-}, z^{+})) = 1-2\alpha$.



$\square$

\end{document}